\documentclass[epj]{webofc}
\usepackage[varg]{txfonts}   % Web of Conferences font
\usepackage{multirow}
\wocname{EPJ Web of Conferences}
\woctitle{ICNFP 2016}
\begin{document}
\selectlanguage{english}
\title{The proton radius puzzle}
%
% subtitle (optional, strongly discouraged)
%
%%%\subtitle{Do you have a subtitle?\\ If so, write it here}

\author{Maurizio Bonesini\inst{1} \fnsep\thanks{\email{ maurizio.bonesini@mib.infn.it}  } 
} 

\institute{Sezione INFN Milano Bicocca, Dipartimento di Fisica G. Occhialini, Milano, Italy \\ (on behalf of the {\bf FAMU Collaboration)} 
}

\abstract{
The FAMU ({\bf F}isica degli {\bf A}tomi {\bf Mu}onici) experiment has the goal
to measure precisely the proton Zemach radius, thus contributing to the solution
of the so-called proton radius ``puzzle''. To this aim, it makes use 
of a high-intensity pulsed muon beam at RIKEN-RAL impinging on a cryogenic
hydrogen target with an high-Z gas admixture  and a tunable mid-IR high power 
laser, to measure the hyperfine (HFS) splitting of the 1S state of the muonic 
hydrogen. From the value of the exciting laser frequency, the
energy of the HFS transition
may be derived with high precision ($\sim 10^{-5}$) 
and thus,  via QED calculations, the
Zemach radius of the proton.
 The experimental apparatus includes a precise fiber-SiPMT beam
hodoscope and  a crown of eight $LaBr_3$ crystals and a few HPGe detectors 
for detection
of the emitted characteristic X-rays. Preliminary runs to optimize the gas
target filling and its operating conditions have been taken in 2014
and 2015-2016. The final run, with the pump laser to drive the HFS transition, 
is expected in 2018. 
}
\maketitle
\section{Introduction: the proton radius puzzle}
\label{intro}
Many properties of the proton - the primary visible universe building block-
such as its radius and anomalous magnetic moment, are not completely
 understood.
The so-called proton radius ``puzzle'' \cite{Antognini} refers to the $7 \sigma$ discrepancy
between the electron and muon determination of the proton charge radius. 
This discrepancy may be due to a violation of the electron-muon universality
or simply to not well understood experimental problems.

The precise measurement of the Zemach radius of the proton with
muons may shed new light on the problem.
Table \ref{tab1} resumes the current experimental situation for both the 
proton charge  ($r_{ch}$)  and Zemach radius ($r_{Z}$)  
determination with muons or electrons. 
\begin{table}[h]
\centering
\caption{The present situation of the proton charge and Zemach radius measurements.}
\label{tab1}       % Give a unique label
% For LaTeX tables you can use
\begin{tabular}{|l|l|l|}
\hline
                                & charge radius $r_{ch}$(fm) & Zemach radius 
$r_{Z}$ (fm) \\ \hline
\multirow{2}{*}{e-p scattering and hydrogen spectroscopy}           & &
            $r_Z=1.037(16)$ \cite{Dupays} (2003)\\ 
&\multirow{2}{*}{$<r_{ch}>=0.8751(61)$ \cite{xx}}             & $r_{Z}=1.086(12)$ \cite{Friar} (2004)\\ 
  &   & $r_Z=1.047(16)$
\cite{Volotka} (2005)\\  
  &   & $r_Z=1.045(4)$ \cite{Distler} (2011) \\
\hline
$\mu^{-}$-p scattering and & $<r_{ch}>=0.84087(39)$ \cite{Antognini}& $r_{Z}=1.082(37) $
\cite{Antognini} \\ 
and Lamb Shift spectroscopy   &               & \\ \hline
\end{tabular}
\end{table}
\section{The FAMU experimental method}
\label{method}
\begin{figure}[hbt]
\centering
\vskip -1cm
\includegraphics[width=0.80\linewidth]{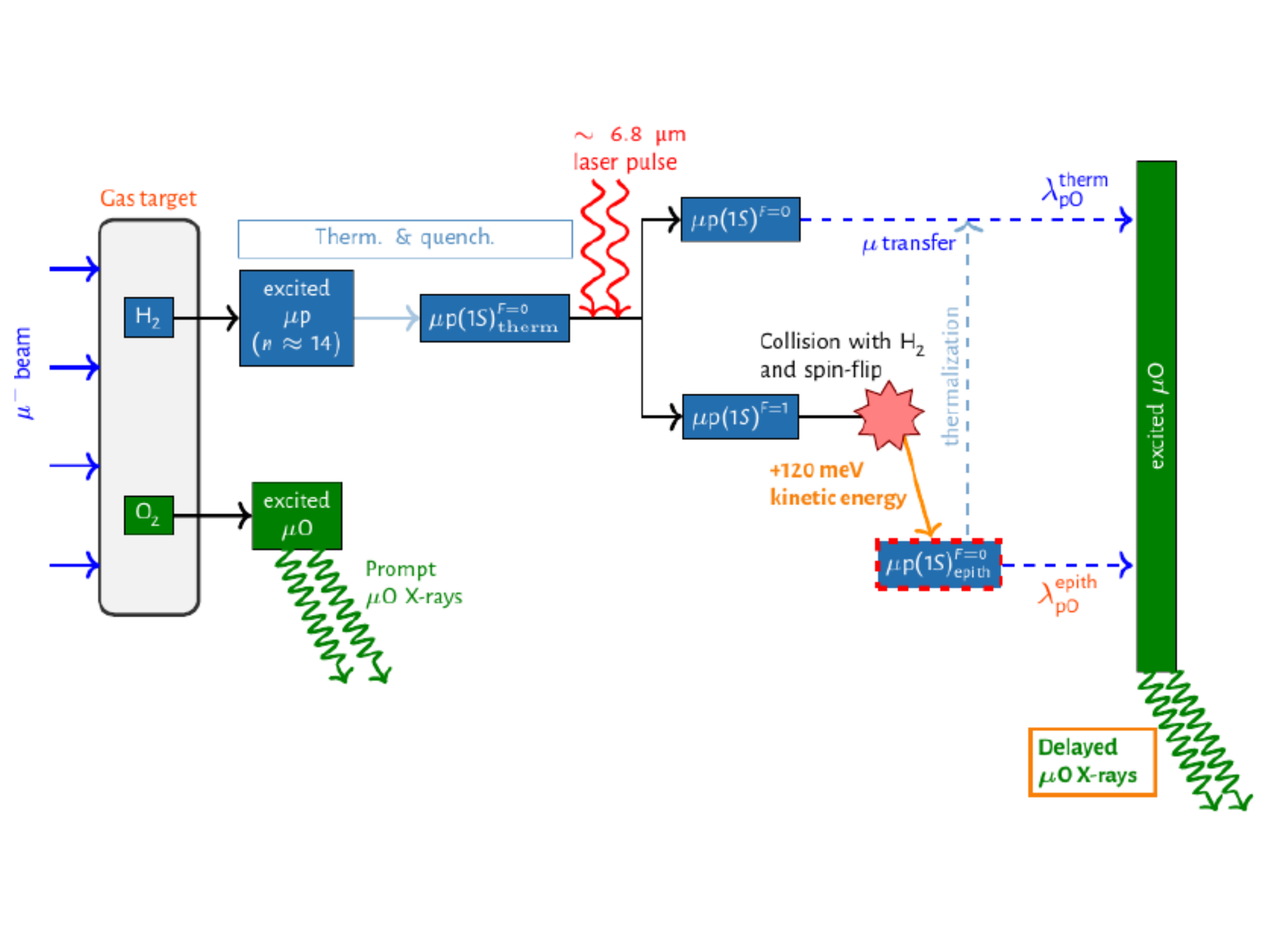}
\vskip -1.0cm
\caption{Schematic illustration of the principle of the FAMU experiment.
After the initial phase of thermalization, a laser pulse will induce the singlet to triplet transitions in the muonic hidrogen atoms. Further collisions with 
$H_2$ molecules will convert back F=1 states to F=0 states with $\sim 0.2 $ keV
kinetic energy ($\sim \Delta E^{HFS}$). Finally, the muon is transferred from 
$(\mu^{-}p)$ to the higher Z admixture gas and the characteristic X-ray 
K-lines from $(\mu Z^{*})$ cascade are observed. }
\label{fig0}       % Give a unique label
\end{figure}

The experimental method to be used by the FAMU Collaboration \cite{bakalov}
is shown schematically  in figure \ref{fig0}. It makes
use of a high intensity pulsed low-energy muon beam, stopping in a hydrogen 
target, to produce muonic hydrogen (in a mixture of singlet F=0 and triplet
F=1 states) and a tunable mid-IR high power laser 
to excite the hyperfine splitting (HFS) transition of the 1S muonic 
hydrogen (from F=0 to F=1 states).
Exploiting the muon transfer from muonic hydrogen to another
higher-Z gas in the target (such as $O_2$ or Ar), 
the $\mu^{-}p_{1S}$ HFS transition will be recognized by 
an increase of the number of 
 X-rays from the $(\mu Z^{*})$ cascade, while tuning the laser
frequency $\nu_{0}$ ($\Delta E_{HFS}=h \nu_{0}^{res}$). 

One can thus measure $\Delta E_{HFS}$ with a relative accuracy $\sim 10^{-5}$ and
determine, via QED calculations, the proton Zemach radius with high
precision.

The FAMU experiment will be performed in steps, starting from the study of
the transfer rate from muonic hydrogen to another higher-Z gas (2014-2016) 
and ending with a full working setup (2018 $\mapsto ... $
) with the pump laser and a cavity to
determine the proton Zemach radius. The preliminary steps will allow
determining the best mixture to be used inside the cryogenic target and 
optimize the operating conditions. 
\section{The experimental setup}
\label{setup}
\begin{figure}[h]
\centering
\includegraphics[width=0.42\linewidth]{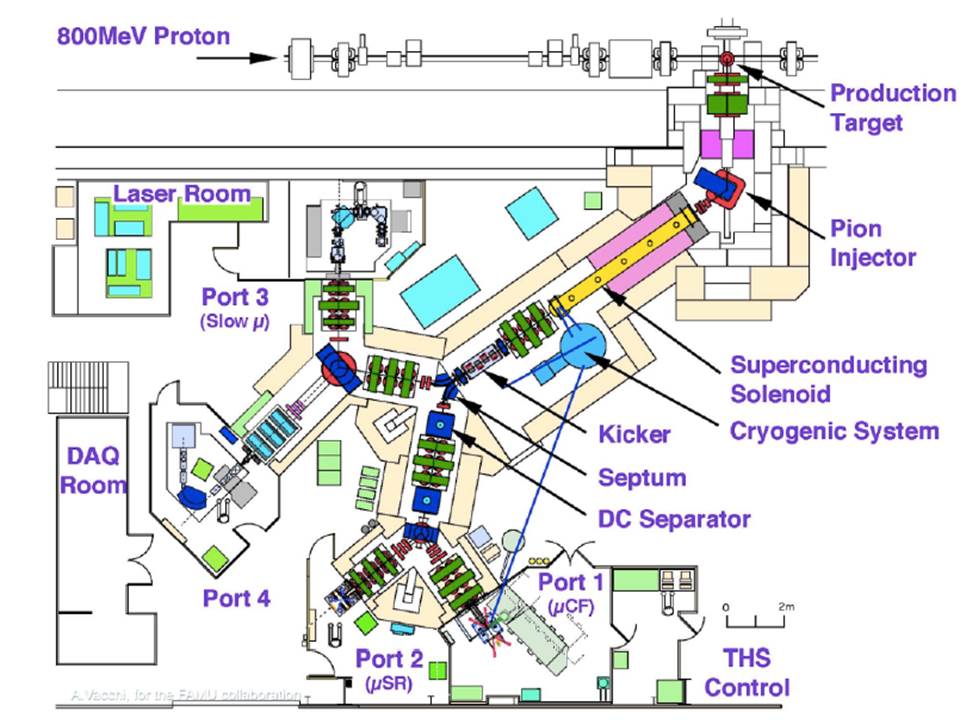}
\includegraphics[width=0.4\linewidth]{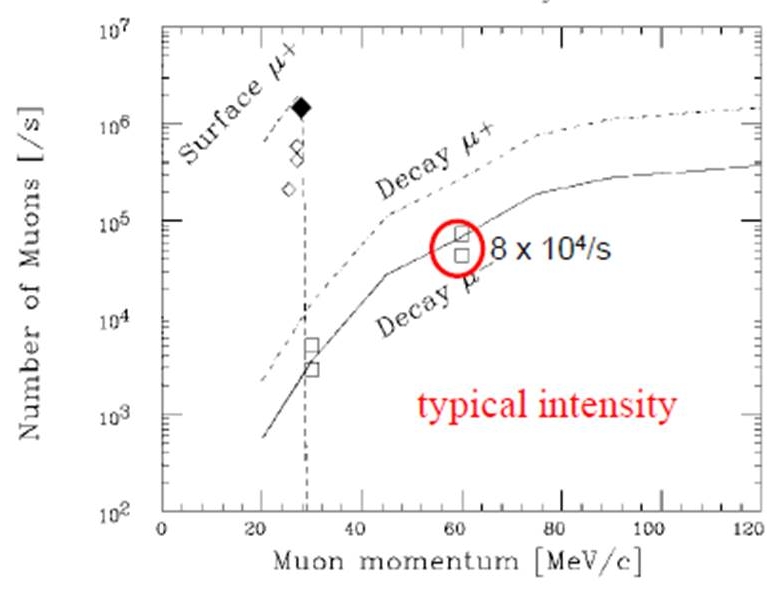}
\caption{Left panel: layout of the RIKEN-RAL facility at RAL, with its
4 experimental ports. The FAMU 
esperiment presently uses PORT 4 and will move to PORT 1 for the final run.
Right panel: estimated intensity of the RIKEN-RAL muon beam for surface muons 
and decay muons in a typical square $4 \times 4$ cm$^2$ area.}
\label{fig1-2}       % Give a unique label
\end{figure}

%%\subsection{The RiKEN RAL muon facility}
%%\label{sec-1}
The RIKEN-RAL muon facility is located at the Rutherford Appleton
 laboratory at Didcot (UK) 
and provides high intensity pulsed muon beams at four experimental 
ports.
Its schematic layout is shown in the left panel of figure \ref{fig1-2}.
The ISIS primary proton beam at 800 MeV/c, with a 50 Hz repetition rate,
impinges from the left on a secondary carbon target producing a high-intensity 
pulsed muon beam, that reflects the primary beam structure: a double pulse
structure with 70 ns pulse width (FWHM) and 320 ns peak to peak distance.
The FAMU experiment makes use of the  decay muon beam (muons produced outside
the target from pion decay and collected by a decay solenoid) at $\sim 60$
 MeV/c. The intensity is around $8 \times 10^4 \mu^{-}/s$, as shown 
in figure \ref{fig1-2}, with an energy spread
$\sim 10 \%$ and an angular divergence $\sim 60$ mrad.
\subsection{The 2014 and  2015-2016 preliminary runs}
\label{sec-3}
\begin{figure}[h]
\centering
\includegraphics[width=0.4\linewidth]{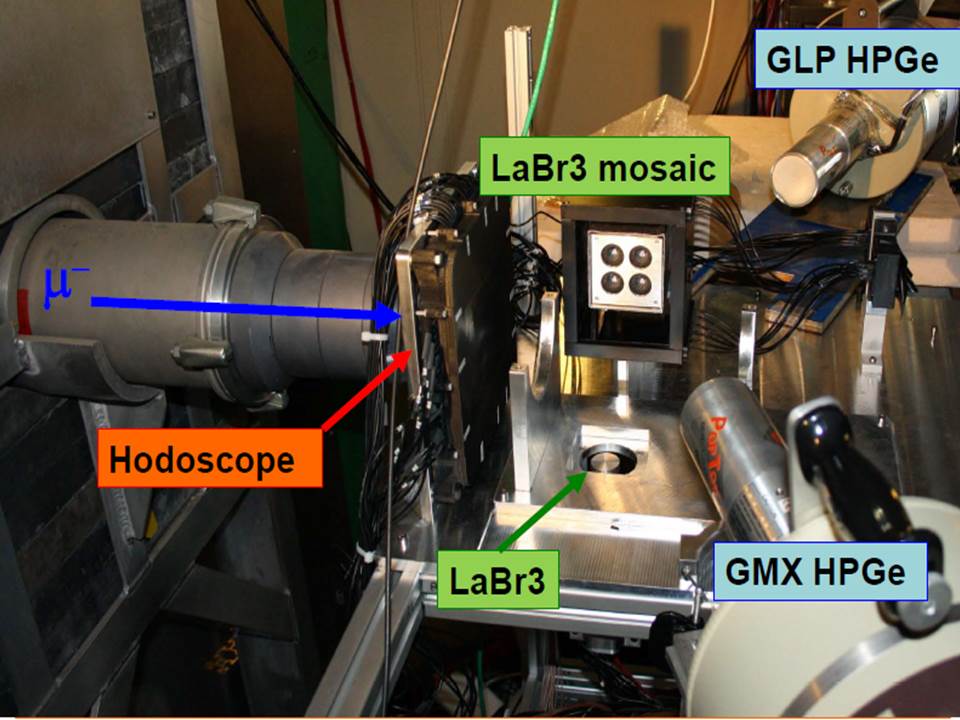}
\caption{Preliminary setup for the 2014 FAMU run at RiKEN-RAL. From left to
right, along the beamline, the 3mm pitch beam hodoscope, the target surrounded 
by two LaBr3 detectors and two HpGe detectors}
\label{fig5}       % Give a unique label
\end{figure}

\begin{figure}[h]
\centering
\includegraphics[width=0.43\linewidth]{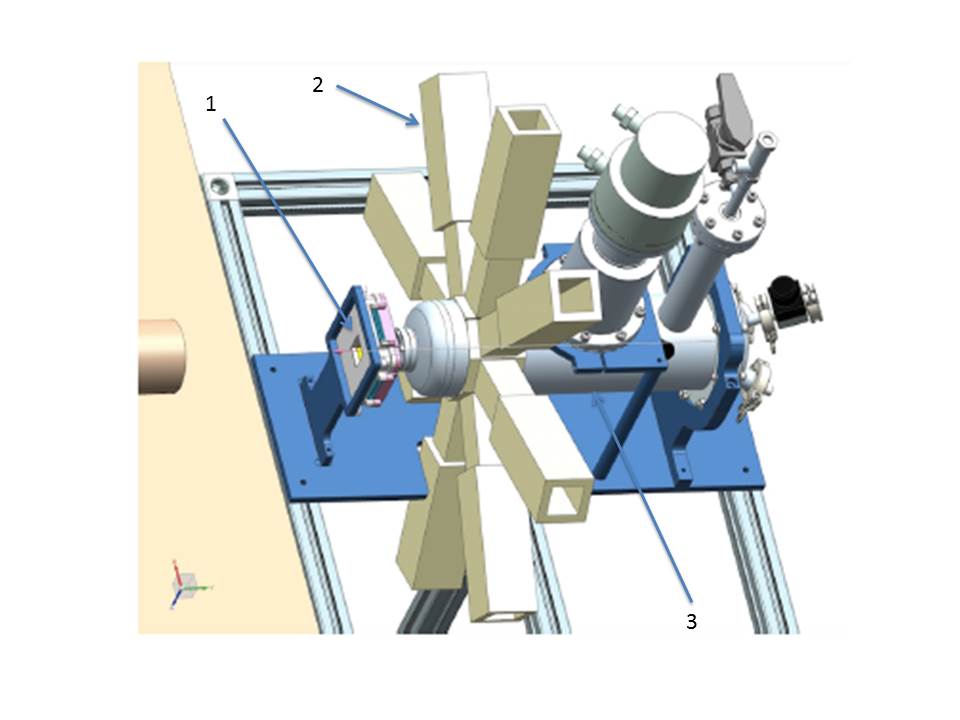}
\includegraphics[width=0.40\linewidth]{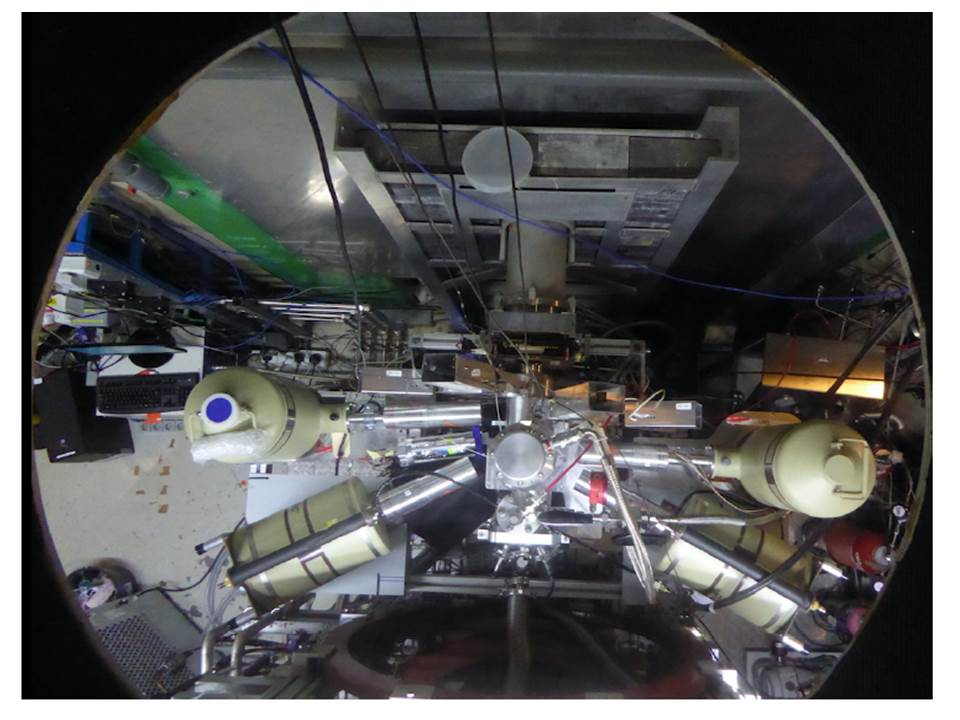}
\caption{Left panel: layout of the FAMU experimental setup for the
2015-2016 run at RIKEN-RAL: (1) is the 1 mm pitch 
beam hodoscope; (2) is the crown of eight LaBr3 crystals with PMT readout, for
fast X-rays detection; (3) is the cryogenic target. Right panel: view from the
top of the experimental setup: the four HpGe detectors, for precise X-rays detection, are also shown. }
\label{fig6-7}       % Give a unique label
\end{figure}

The setup for the 2014 (R484) and 2015-2016 (R582) 
FAMU data taking are schematically shown
in figures \ref{fig5} and \ref{fig6-7}. 
A fiber-SiPMT 
beam hodoscope is the most upstream detector of the setup and had
a 3 mm pitch in the first 2014 run and  a 1 mm pitch in the following ones. 
It was used to optimize beam steering inside the target.
Both beam hodoscopes (with 1 mm and 3 mm pitch) have a similar structure based 
on BCF12 Bicron square single-clad 
scintillating fibers along X/Y directions (32+32 channels)
read at one end by Advansid SiPMT. Difficulties encountered in 
developing the mechanics of the two beam hodoscopes (especially the one with
1 mm pitch) were solved via  3D printing.
Details of the construction and preliminary performances of the first beam
hodoscope were reported in references \cite{famu} and  \cite{carbone}. 

\begin{figure}[h]
\centering
\includegraphics[width=0.40\linewidth]{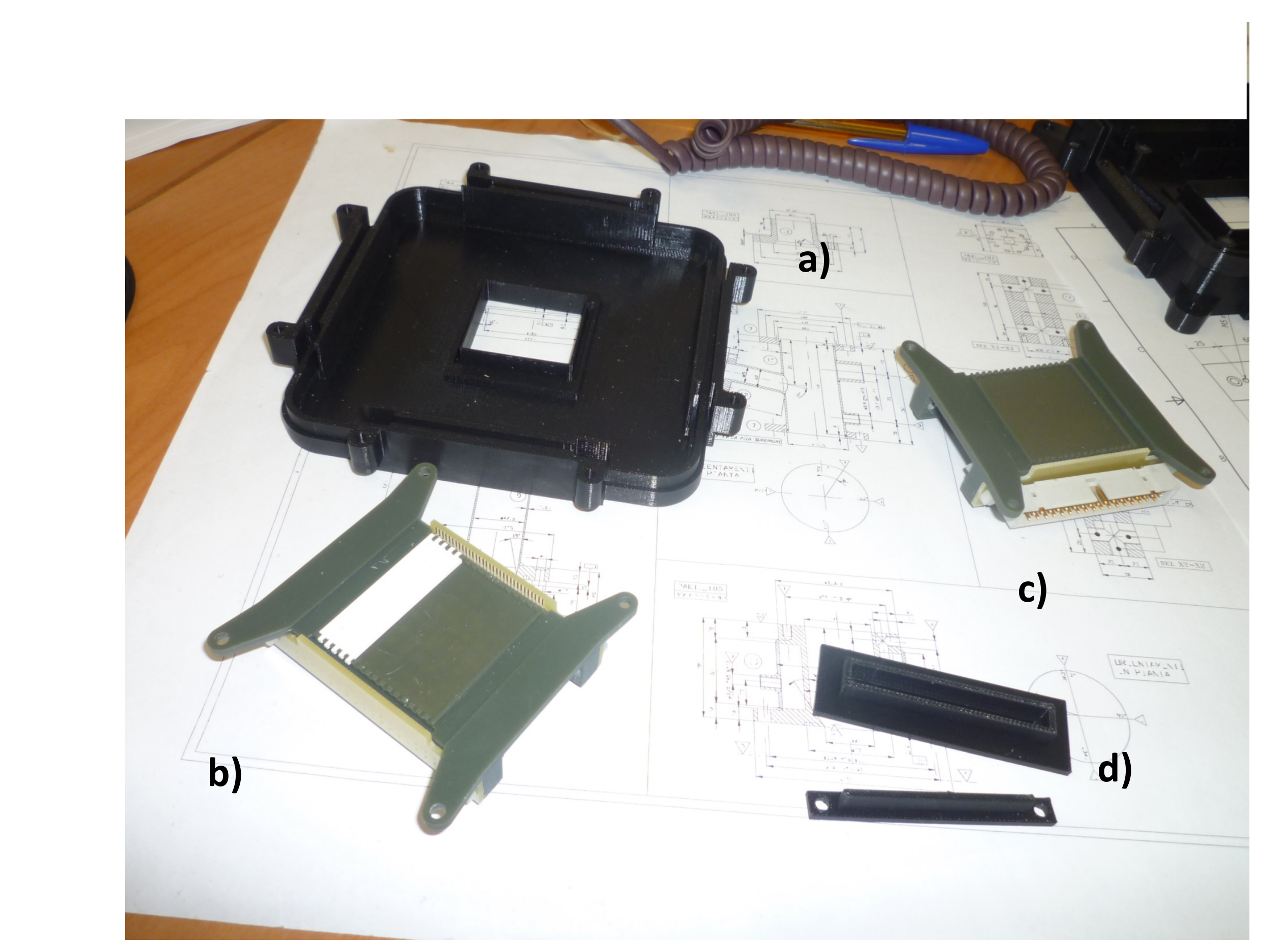}
\includegraphics[width=0.25\linewidth]{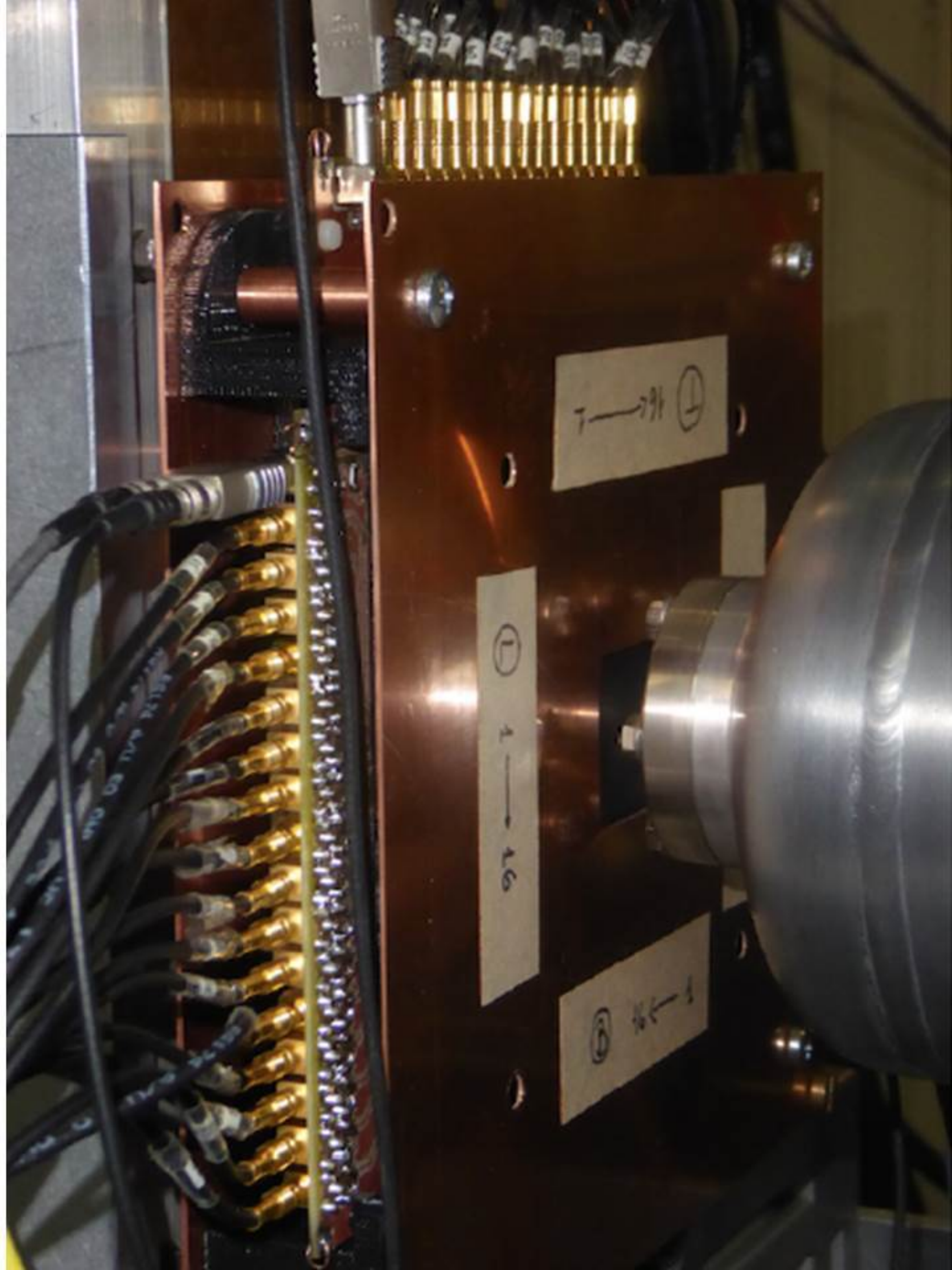}
\vskip -0.3cm
\caption{Left panel: main mechanical components of the 1mm pitch beam hodoscope:
(a) one of the two light tight beam hodoscope cover boxes, 
(b),(c) X/Y plane 
fiber holders: in (b) some fibers are mounted, while 
in (c) the flat cable connector is shown , (d) cable feed-through. Right panel: 
picture of the beam hodoscope after the beam collimator}
\label{fig8}       % Give a unique label
\end{figure}

The 1 mm pitch beam hodoscope (active area $\sim 3.2 \times 3.2$ cm$^2$)
was developed to be put in front of the thin Beryllium window of the 
cryogenic target, used in the 2015-2016 run. An effort was done to minimize
materials present in the beamline. Thus Bicron BCF12 square $1 \time 1$ mm$^2$ 
scintillating fibers, with white EMA coating to avoid light cross-talk, were
used. 
Scintillating fibers were cut at Cern with a Fiberfin4 machine, that produced
ready-to-use fibers of the right length with polished ends.
The main mechanics components and a picture of 
the detector after the beam collimator are shown in figure 
\ref{fig8}. 

\begin{figure}[h]
\centering
\vskip -1.5cm
\includegraphics[width=0.35\linewidth]{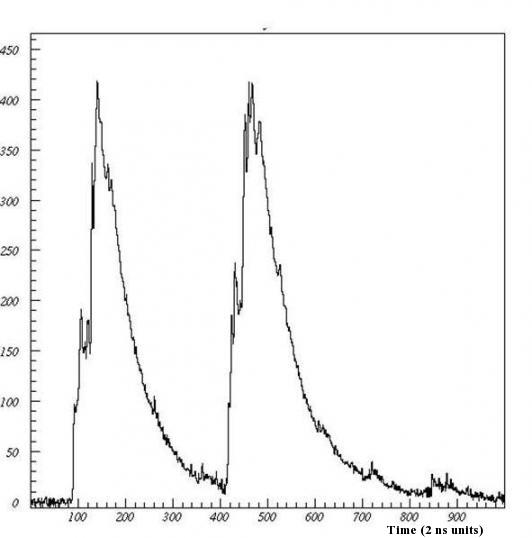}
\includegraphics[width=0.57\linewidth]{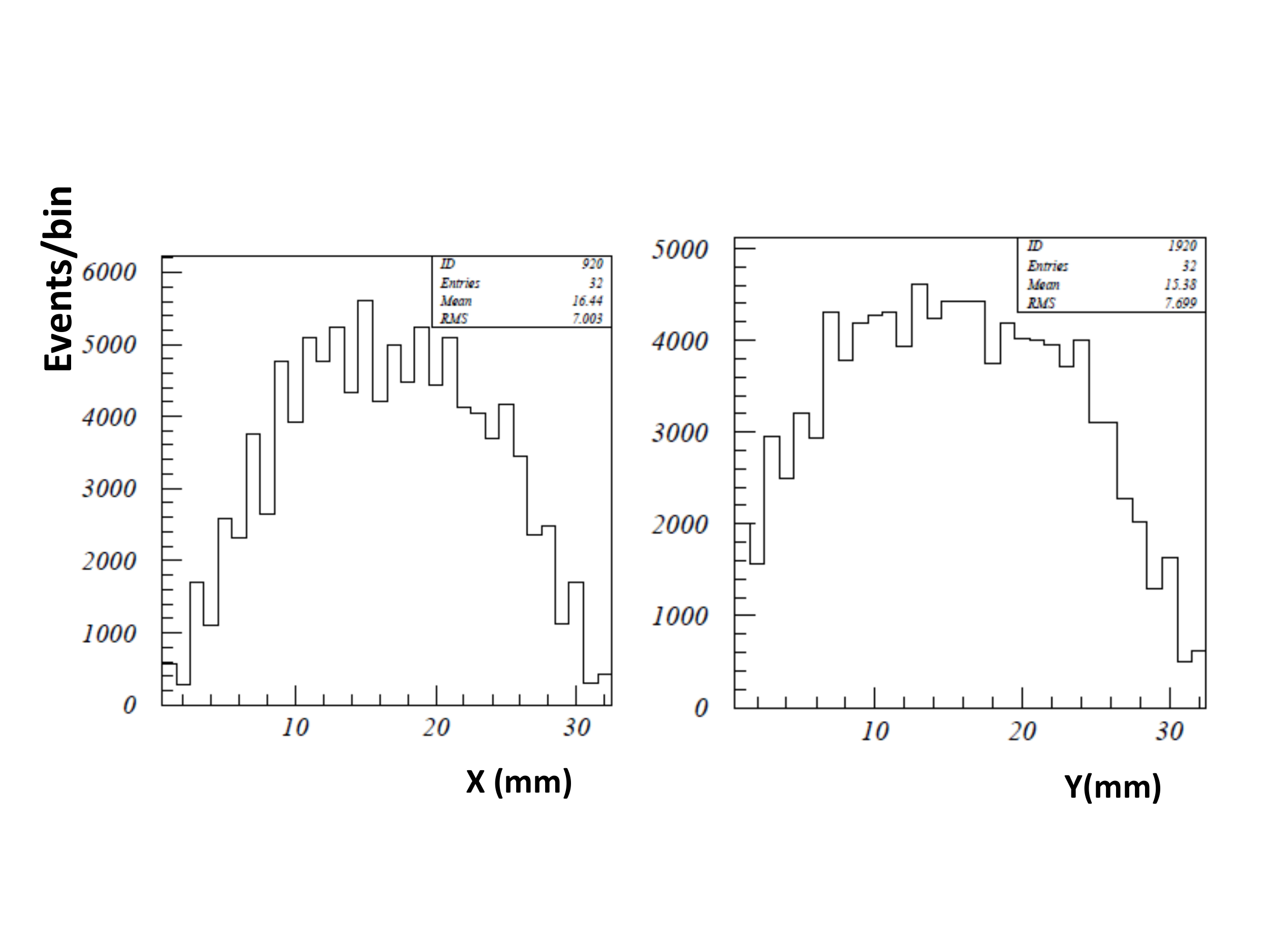}
\caption{Left panel: waveform, as recorded by  V1742 FADCs, for a typical
channel of the 1 mm pitch beam hodoscope. Right panel: x/y beam profile,
with standard optics, as recorded by integrating the waveform signals
during the 2016 data taking. }
\label{fig_ab}       % Give a unique label
\end{figure}

Scintillating fibers were read from one edge only, alternating
 up/down or left/right sides, by RGB $1\times 1$ mm$^2$ 
Advansid SiPMT with $40 \ \mu$m cells. As the breakdown voltages ($\sim 29$ V) 
were very similar for all SiPMTs:
 within 100 mV, it was possible to use a simplified front-end 
electronics with respect to the one of the previous 3 mm pitch beam 
hodoscope, derived from the INFN TPS project~\footnote{each TPS electronic
board, with 8 individual channels, provided individual SiPMT voltage 
regulation, signal amplification and shaping and signal discrimination and
trigger capability. Individual output signals were then fed into a CAEN 
V792 QADC, to measure the charge integrated signal}.
All channels of the same plane (x,y) of the 1mm pitch beam hodoscope  were
instead  powered by a single
HV channel and the output signals were directly fed into a fast FADC~\footnote{
CAEN V1742 FADC with 5 Gs/s, 12 bit, 1 Vpp dynamic range  in VME standard}. 
Signals were routed from the PCB's, where the SiPMTs were mounted, to the 
external MCX cables, going to the FADC, via a short flat cable. The impedance
mismatch of this cable was tested and found negligible. 
Figure \ref{fig_ab} shows the FADC waveform for a typical channel, where the
two pulses beam structure is evident (left panel) and the reconstructed x,y 
beam profiles from the integrated charge information. 
As expected, the beam profile is dominated by the size of the collimator 
and has an RMS $\sim 7-8$ mm. 

\begin{figure}[h]
\centering
\includegraphics[width=0.55\linewidth]{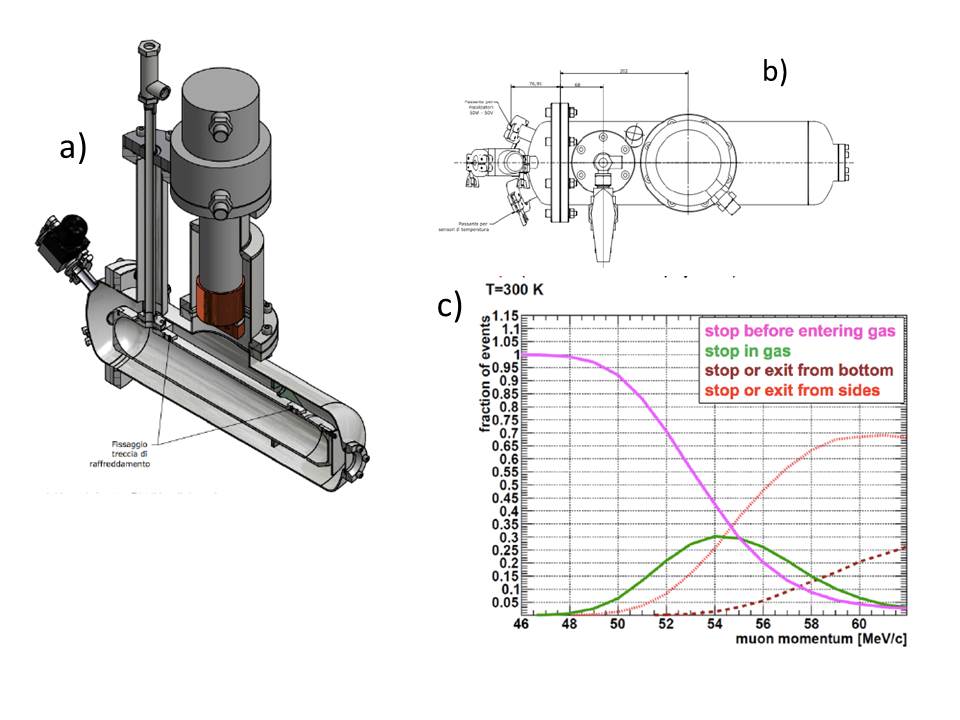}
\includegraphics[width=0.44\linewidth]{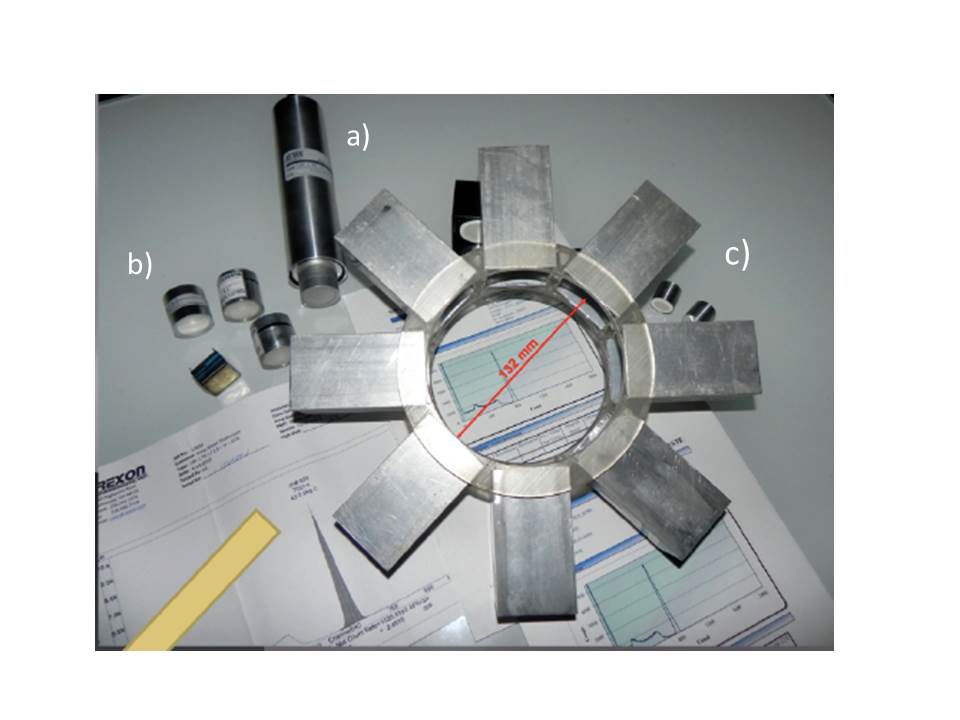}
\vskip -0.5cm
\caption{Right panel: (a) layout of the cryogenic target; (b) top view of the 
cryogenic target;
(c) stopping of muons inside the target at T=300K.
Left panel: a picture of the eight 1" LaBr3(Ce) detectors arranged in a star.
(a) Hamamatsu R11265-200 UBA PMTs; (b) 1" LaBr3(Ce)
crystals; (c) mechanics arrangement}
\label{fig9}       % Give a unique label
\end{figure}

The target, used in the 2014 run, 
 was an Al cylindrical vessel~\footnote{with a 125 mm bare diameter
and a 260 mm length, it had a 2.8 liters volume. The walls had a thickness
of 7 mm, except the entrance window thinned to 4 mm.} 
filled with a high-pressure gas. In the following runs, it was replaced 
by a cryogenic
one. The layout of the cryogenic target is shown in the left panel of figure 
\ref{fig9}, where the percentage of stopped muons as a function of the beam momentum is shown nearby. 
Emitted X-rays are detected with  high precision Germanium detectors (HpGe)
and fast LaBr3:Ce detectors.
Two HPGe detectors (one small area ORTEC GLP detector and one ORTEC GMX 
detector) 
were used in the 2014 run, while in the following runs two additional HPGe
detectors from RIKEN-RAL were also used.
In the 2014 run a mosaic of four 0.5-inch diameter, 0.5-inch length 
$LaBr_3 (5\% Ce)$ read by Hamamatsu R11265-200 UBA photomultipliers and
a single 1-inch diameter, 1-inch length Brillance 380 $LaBr_3$ crystal
read by a Phillips
XP2060 photomultiplier were used. 
In the following runs, the $LaBr_3$ mosaic was replaced by a crown of eight
1-inch 
$LaBr_3$ detectors read by Hamamatsu R11265-200 UBA photomultipliers, 
with an active divider, as shown in figure \ref{fig9}.
The $LaBr_3$ detectors were read through 500 Mhz, 14-bit digitizer (CAEN 
V1730) with a $5 \mu$s window after the trigger, provided by the RF of 
the beamline, while for the HPGe detectors a slower 100 MHz, 14-bit
 digitizer 
(CAEN V1724) was used. 
In addition, during the 2016 data new X-rays detectors, developed for PET,
 based on PrLuAG 
and CeCAAG crystals and read out by Hamamatsu 
TSV SiPMT arrays, were introduced
to instrument the volume under the cryogenic target \cite{eps15}. 

\subsection{The setup for the final step of the FAMU experiment.}
\label{final}
\begin{figure}[h]
\centering
\includegraphics[width=0.7\linewidth]{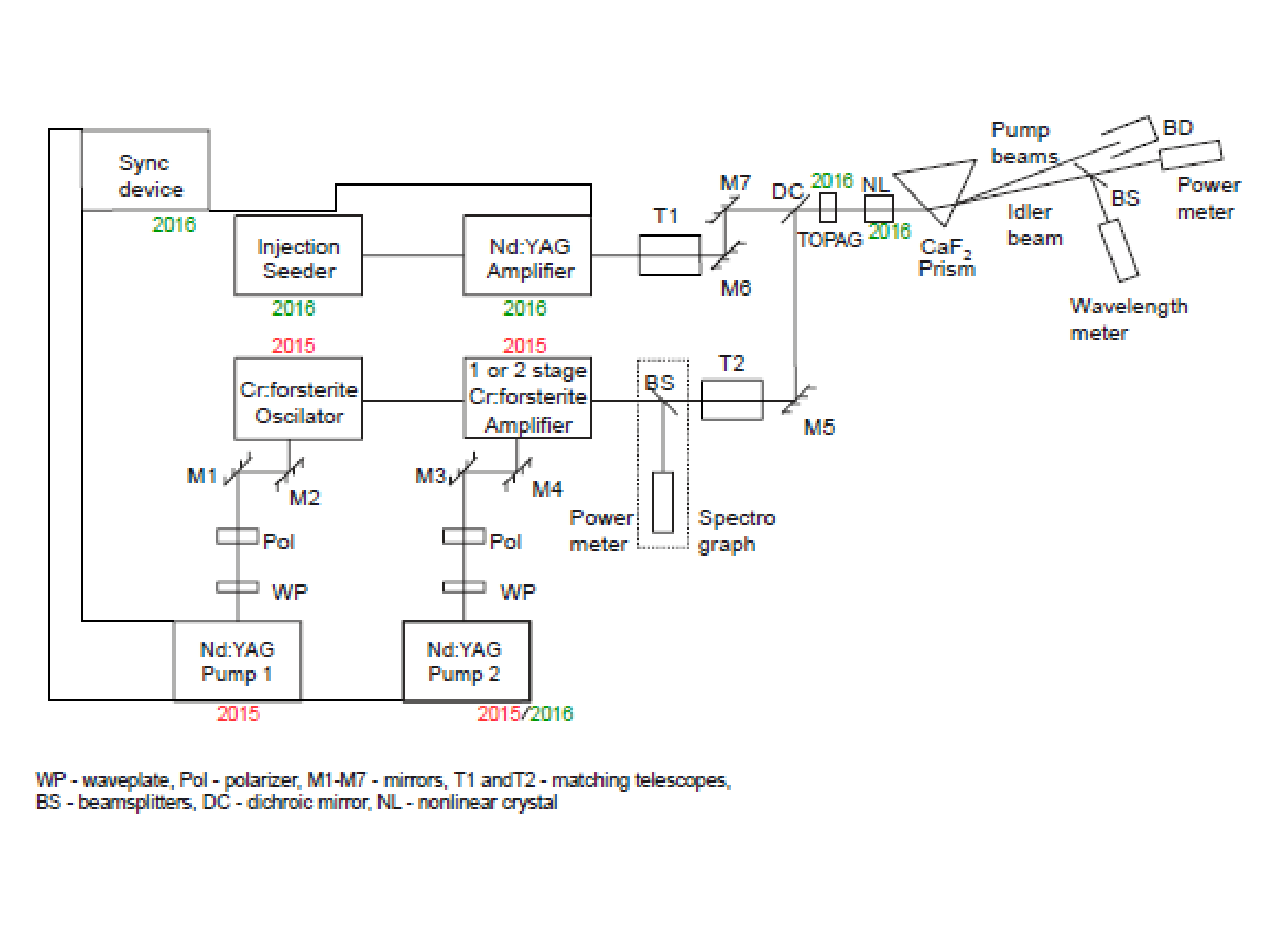}
\vskip -1cm
\caption{ (Structure of the proposed mid-IR laser, under development
at Electra (Ts). Wp: waveplate, Pol: polarizer; M1-M7: mirrors; T1-T2:
matching telescopes; BS: beamsplitter; DC: dichroic mirror; NL: nonlinear
crystal.}
\label{fig15}       % Give a unique label
\end{figure}

To excite the HFS transition, a laser tunable at a wavelength 
around $\lambda_{0} \sim 
\frac{2 \pi h c}{\Delta E_{HFS}} \sim 6.785 \ \mu$m will be used.
In reference \cite{vacchi}, the main requirements on the laser system,
such as energy ($\ge 0.25$ mJ) and linewidth precision ($\lambda_{0}
\leq 0.07$ nm )
are discussed.
The layout of the proposed laser system, under development at the Electra 
laboratory (Ts), is schematically shown in 
figure \ref{fig15}. A scheme based on direct difference frequency
generation in nonlinear crystals, using a Q-switched Nd-Yag laser and
a tunable Cr:forsterite laser will be used and is presently under test
\cite{lubo}. 
 
To increase the probability of the spin-flip (F=0 to F=1) 
process either the target 
temperature may be lowered or a multipass cavity may be used: see
reference \cite{Vogelsang} for an example. With a multipass
cavity that provides $\sim 2 \times 10^3$ reflections, the spin-flip
probability would reach a reasonable $\sim 12 \%$ value, that makes the 
experiment feasible. Studies of a suitable multipass cavity, that
may accomodate the available cryogenic target,  are under way inside
the FAMU collaboration. 

\section{Detector performances and preliminary results}
\label{detector}
\begin{figure}[h]
\centering
\includegraphics[width=0.45\linewidth]{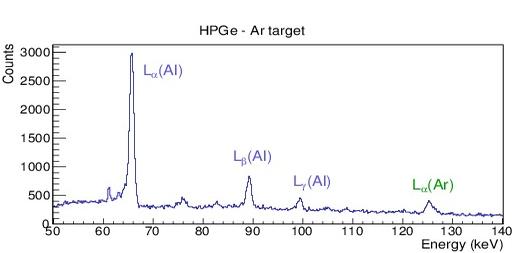}
\includegraphics[width=0.51\linewidth]{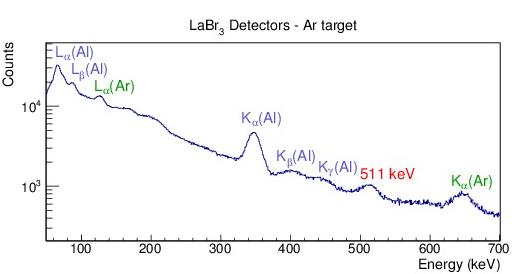}
\vskip -0.1 cm
\caption{Left panel: characteristic X-rays lines, as detected by the HPGe 
detectors. Right panel: the same as detected by the $LaBr_3$ detectors.}
\label{fig11}       % Give a unique label
\end{figure}

The aim of the 2014 run (R482) was to test the suitability of the proposed
setup for the foreseen measurement: in particular the detection of the
characteristic X-rays in a large background environment. The detection 
of characteristic X-rays from $LaBr_3$ detectors was cross-checked with 
two high precision HpGe detectors.   
As an example, for an $H_2+ 2\% Ar$ filled target, figure \ref{fig11} 
shows the spectra detected by the HPGe detectors and the $LaBr_3$ 
crystals. 
All characteristic X-ray lines from the Ar admixture are detected, showing
the correctness of the method.

\begin{figure}[h]
\centering
\vskip -1cm
\includegraphics[width=0.70\linewidth]{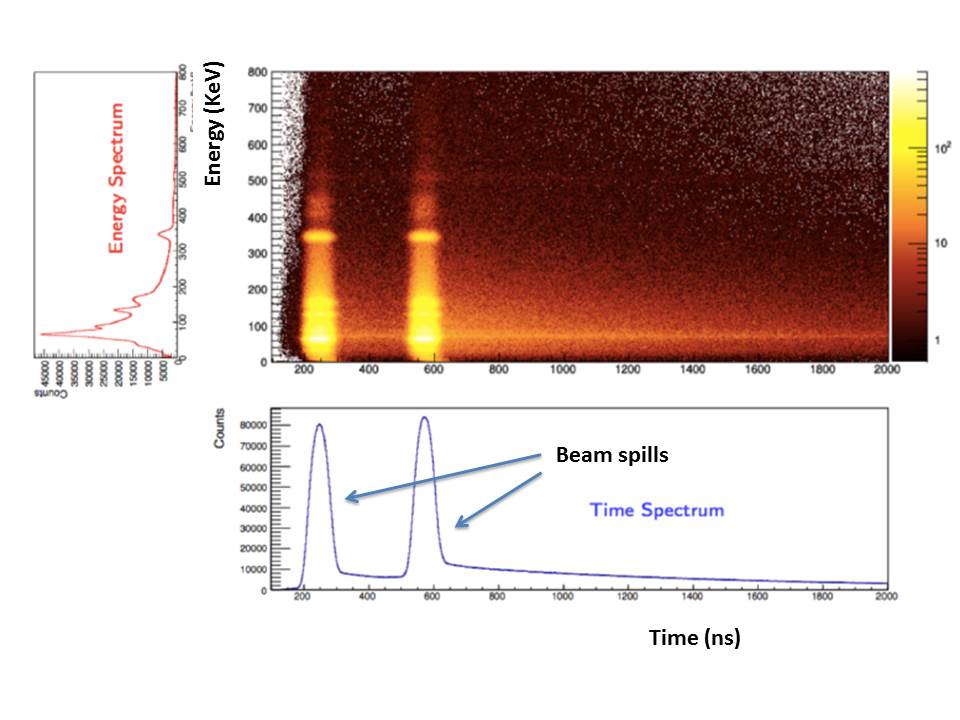}
\vskip -1.0 cm
\caption{A snapshot of the detected X-ray spectrum (energy vs time). The
two peaks are due to the two pulses structure in the incoming muon beam.}
\label{fig16}       % Give a unique label
\end{figure}

Regarding the time distribution of the events, figure \ref{fig16} shows an
example of the time evolution of the X-ray spectrum for the $H_2-CO_2$ target.
The prompt peaks, corresponding to the two beam spills, are followed by a tail
populated by the products of muon decay. From the events in the tails, 
the lifetime of muonic atoms may be estimated. Results are reported in table
\ref{tab-d} and are compatible with previous ones, as reported in 
reference \cite{Suzuki}.   

\begin{table}[h]
\centering
\caption{Lifetimes of some muonic atoms in ns, as determined in the FAMU 
2014 run.}
\label{tab-d}       % Give a unique label
% For LaTeX tables you can use
\begin{tabular}{|c|c|c|}
\hline
        & FAMU~\cite{famu} (ns) & T. Suzuki {\it et al.} \cite{Suzuki} and references \\ 
        &                & therein (ns) \\ \hline
$\mu$C  & $2011 \pm 16 $ & $2026 \pm 1.5$ \\
$\mu$Al & $879 \pm 28 $  & $864 \pm 2$ \\
$\mu$O  & $1824 \pm 46$  & $1795 \pm 2 $ \\
$\mu$Ar & $564 \pm 14 $  & $537 \pm 32 $ \\ 
$\mu$p  & $2141 \pm 98$  & $2194.53 \pm 0.11$ \\
\hline
\end{tabular}
\end{table}
Studies are under way to determine the transfer rate from hydrogen to
the gas mixture contaminants (Argon or Oxygen), using the time evolution
of the X-rays events inside the Argon (Oxygen) peak. 
 \section{Conclusions}
The feasibility of the method proposed by the FAMU Collaboration to 
determine the proton Zemach radius has been demonstrated with preliminary
measurements at RIKEN-RAL: in particular the possibility to use a high-rate
X-rays detection system based on $LaBr_3$ crystals.

The mid-IR pulsed laser and the multipass cavity are under development, 
showing that a final data taking in 2018 is feasible.

\end{document}